\documentclass[aip,jcp,reprint,superscriptaddress]{revtex4-1}

\usepackage[utf8]{inputenc}
\usepackage[T1]{fontenc}
\usepackage[english]{babel}
\usepackage[dvips]{graphicx}
\usepackage{fullpage}
\usepackage{lmodern}
\usepackage{float}
\usepackage{listings}
\usepackage{color}
\usepackage{bold-extra}
\usepackage[colorlinks,breaklinks,linkcolor=blue,citecolor=blue,urlcolor=blue]{hyperref}

\usepackage{amsmath}
\usepackage{amssymb}
\usepackage{amsfonts}
\usepackage{braket}

\newcommand{\rv}{\textbf{r}}

\newcommand{\del}{\nabla}

\newcommand{\td}{\text{d}}
\usepackage{varwidth}

\begin{document}

\title{Orbital-free density functional theory implementation with the projector augmented-wave method}

\author{Jouko Lehtom\"aki}
\affiliation{COMP Centre of Excellence, Department of Applied Physics, Aalto University, P.O. Box 11100, 00076 Aalto, Finland}
\author{Ilja Makkonen}
\affiliation{COMP Centre of Excellence, Department of Applied Physics, Aalto University, P.O. Box 11100, 00076 Aalto, Finland}
\author{Miguel A. Caro}
\affiliation{COMP Centre of Excellence, Department of Applied Physics, Aalto University, P.O. Box 11100, 00076 Aalto, Finland}
\affiliation{Department of Electrical Engineering and Automation, Aalto University, Espoo, Finland}
\author{Ari Harju}
\affiliation{COMP Centre of Excellence, Department of Applied Physics, Aalto University, P.O. Box 11100, 00076 Aalto, Finland}
\author{Olga Lopez-Acevedo}
\email{olga.lopez.acevedo@aalto.fi}
\affiliation{COMP Centre of Excellence, Department of Applied Physics, Aalto University, P.O. Box 11100, 00076 Aalto, Finland}

\begin{abstract}
We present a computational scheme for orbital-free density functional theory (OFDFT)
that simultaneously provides access to all-electron values and preserves the OFDFT linear 
scaling as a function of the system size. Using the projector augmented-wave method (PAW) in combination with real-space methods we overcome some obstacles faced by other available implementation schemes. Specifically, the advantages of using the PAW method are two fold. First, PAW reproduces all-electron values offering freedom in adjusting the convergence parameters and  the atomic setups allow tuning the numerical accuracy per element. Second, PAW can provide a solution to some of the convergence problems exhibited in other OFDFT implementations based on Kohn-Sham codes. Using PAW and real-space methods, our orbital-free results agree with the reference
all-electron values with a mean absolute error of 10~meV and the number of
iterations required by the self-consistent cycle is comparable to the
KS method. 
The comparison of all-electron and pseudopotential bulk modulus and lattice constant reveal an enormous difference, 
demonstrating that in order to assess the performance of OFDFT functionals it is necessary to use implementations that obtain all-electron values. The proposed combination of methods is the most promising route currently available. We finally show that a parametrized kinetic energy functional can give lattice constants and bulk moduli comparable in  accuracy to those obtained by the KS PBE method, exemplified with the case of diamond. 
\end{abstract}

\maketitle

\section{Introduction}

The most popular implementation of density functional theory (DFT), in
principle an exact theory,~\cite{HK} has been the Kohn-Sham (KS) method,
which offers accuracy at reasonable computational cost.~\cite{KS}
Orbital-free density functional theory (OFDFT) is another calculation
scheme that in principle offers better scaling, but has not been the
focus of intense research thus far. 
In OFDFT theory, only functionals  that are explicitly dependent on the total
density are used, making it theoretically capable of scaling to much larger
systems than KS. 
The solution of the OFDFT problem can also be cast in a form which is easy to
implement reusing existing KS codes.~\cite{orbital_free} Within this approach,
one solves a KS-like differential equation for only one orbital 
describing the full system. 
In addition, when only one orbital is present in a KS-like calculation,
it is also possible to achieve better performance in orthonormalization
and subspace diagonalization,
the computational bottlenecks of current KS DFT codes. Both steps scale in a 
real-space code with state-of-the-art algorithms, for example, as
$\mathcal{O}(N_\text{g} N_\text{b}^2)$, where $N_\text{b}$ is the number of
electronic bands (a number that scales as the number of atoms in the system)
and $N_\text{g}$ is the number of grid 
points (a number that scales as the volume of the simulation cell) in the
calculation.~\cite{performance} Thus, when using the orbital-free model,
we greatly reduce the computational time of these operations by fixing
$N_\text{b}$ to a number that is independent of the system size. 
Moreover, when removing the dependence on the number of bands, the scaling of the overall calculation depends on the number of grid points and the number of atoms. Among the contributions to the overall time scaling arising from the different algorithms employed, the highest-order dependence with the number of grid points or number of atoms is first order. OFDFT can then be implemented as a linear scaling method in a real-space KS DFT code like GPAW~\cite{GPAW}.

Because of the difficulties in convergence and implementation, all-electron
implementations of OFDFT have only been used to derive the energies of
small systems, such as atoms and dimers.~\cite{parr, Karasiev}. On the other hand, OFDFT codes for studying solids rely on the use of
local pseudopotentials (LPPs). The derivation of such pseudopotentials 
requires a parameter fitting, and it has been a common choice to
fit the LPPs to reproduce experimental or KS bulk quantities such as the
lattice parameter, bulk modulus or electronic density profile.~\cite{Carter_review} OFDFT simulations with pseudopotentials have been applied successfully to the study, for example, of the fracture propagation in bulk aluminum or the 
structure factor of warm dense matter.~\cite{Warm_Al} 

As a generalization of pseudopotentials and the linear-augmented-plane wave method the projector-augmented wave (PAW) method has been proposed as a method that allows the calculation of all-electron values with high accuracy and efficiency~\cite{PAW_original}. An {\it{all-electron method}},  refers to methods that solve the Schr\"odinger equation for an electronic system directly in the presence of the $1/r$ non-modified nuclear potential and {\it{all-electron values}} are those values that would be obtained with all-electron methods. In the DFT context, it is standard to term all-electron methods also methods that calculate all-electron values with the use of transformed equations. In this work, we refer to both types of all-electron methods; but to avoid any confusion, the expression {\it{all-electron method}} will refer only to the methods using a non-modified nuclear potential. When referring to the PAW method, we will consistently refer to it as a method obtaining {\it{all-electron values}}. The PAW method works by providing an exact transformation between conveniently smooth pseudo-wave functions and the all-electron wave functions. The PAW transformation provides thus access to all-electron densities, since it is always possible to reconstruct the all-electron wave functions from the pseudo-wave function. Additionally to the PAW transformation, it is also possible to use fixed core orbitals using the frozen core approximation. Moreover, a reliable description over the whole periodic table with good transferability of PAW potentials has been obtained~\cite{GPAW}. As we will demonstrate in the following, using PAW in OFDFT allows the calculation of all-electron energies while also improving the convergence capabilities of the OFDFT implementations reusing KS codes. All-electron values are needed to assess the performance of OFDFT functionals and improve them towards better transferability and accuracy.

The grid-based projector augmented-wave method~\cite{GPAW}
(GPAW) rises above others as a potential platform for OFDFT. It provides
all-electron accuracy, real-space grids and the projector augmented-wave
(PAW) formalism,~\cite{PAW_original} which has proven itself as a viable
method to obtain all-electron values.
As it turns out, the use of the PAW method also helps to stabilize some
of the convergence problems observed earlier in similar OFDFT
calculations (see Ref.~\onlinecite{Karasiev}). Here, we present a novel OFDFT 
implementation in GPAW that simultaneously preserves linear OFDFT scaling,
provides access to all-electron values, and offers improved convergence
capabilities.

The structure of the paper is as follows. After a general introduction to OFDFT equations and PAW transformation, we present results computed with an atomic all-electron code first and then with a real-space PAW code. The atomic all-electron code is used in GPAW to generate the setups (the object containing precomputed values that define the PAW transformation).  With the derived OFDFT setups we calculate, using real-space and PAW methods, binding energies of dimers. Our PAW binding energies agree in the meV range with binding energies from reference all-electron methods. With the OFDFT setups we also compute properties of periodic systems (lattice constant and bulk modulus). Finally, we present some considerations on the performance of the OFDFT implementation using GPAW.

\section{Implementation}

Reusing a KS DFT code can offer important advantages. Currently, KS DFT codes 
are developed by large communities of scientists working
across many countries. Optimization, bug fixes and state-of-the-art methods
are implemented with a quality control that is guaranteed by peer testing
and discussion through open forums.  The present OFDFT implementation will
not only inherit the use of the highly performing algorithms but will also
evolve naturally, being maintained and up-to-date with the evolution of the
underlying KS DFT code. 

\subsection{Orbital-free differential equation}

Our starting point is an exact differential equation for the density and its
natural implementation into a KS code.~\cite{orbital_free} It provides
a Schr\"odinger-like differential equation for a one-orbital model, where
this orbital is the square root of the electronic density. In principle,
this method can be applied for any explicit density functional.

First, the system's total energy functional $E[n]$ (which can be approximate
or exact) is written, using atomic units, in the form
\begin{align}
\label{Evn}
E[n] = & \int \td \rv \, n^{1/2}\left(-\frac{1}{2}\del^2\right)n^{1/2} 
\nonumber \\
& + G[n] + \int \td \rv \, v(\rv) \, n(\rv) 
\nonumber \\
& + \frac{1}{2}\int \td \rv \int \td \rv' \, n(\rv) \, n(\rv ') \,
|\rv - \rv'|^{-1},
\end{align}
where $n(\rv)$ is the electronic density. We immediately recognize the
external potential $v(\rv)$ in the third term and the Hartree potential
in the last term. Here, $G[n]$ is a universal functional that 
that subtracts the first fictitious kinetic energy term and adds the true kinetic energy of the system as well as many-body contributions to the interaction energy not contained in the last Hartree term. The first term, which has the form of a single-particle kinetic energy,
is an auxiliary quantity introduced with the purpose of expressing $E[n]$
in a way that resembles the expression of the KS energy functional. This
will facilitate the implementation of the OFDFT expressions in pre-existing 
KS codes. More precisely, this first term in the energy functional is the kinetic energy of a wave function that equals the square root of the density $n^{1/2}$. It happens to be equivalent to the von Weizs\"acker gradient correction term $T_\text{W} [n]=  \frac{1}{8} \int \td\mathbf{r} \, \frac{\left|\del n(\rv) \right|^2}{n(\rv)}$ (Ref.~\onlinecite{vW}).~\footnote{
Within the present approximation, where the wave function $\phi_o$
equals $n^{1/2}$,
%\begin{align}
%\label{vWTerm}
$\int \td\mathbf{r} \, \phi_0 (\rv) \left( -\frac{1}{2} \del^2 \right) \phi_0(\rv)
= \int \td\mathbf{r} \, \frac{1}{2} \left| \del \phi_0 (\rv) \right|^2
%\nonumber \\
= \frac{1}{8} \int \td\mathbf{r} \, \frac{\left|\del n(\rv) \right|^2}{n(\rv)}
= T_\text{W}[n]$
%\end{align}
In the above equation, the first equality is given by the divergence
theorem. The surface integral associated with the divergence
theorem, $\protect\oint \text{d}\textbf{S} \, \nabla \phi_0^2 (\rv)$, can be
easily shown to vanish: for finite systems $\phi_0 \rightarrow 0$
as $r \rightarrow \infty$; for periodic systems each infinitesimal contribution
to the integral has an equivalent contribution of opposite sign because of
the periodic boundary conditions.
The second equality follows from the chain rule 
$\frac{\partial}{\partial x} \sqrt{n(\rv)} = \frac{\partial n(\rv)}{\partial x}\frac{\partial}{\partial n}\sqrt{n(\rv)} 
= \frac{1}{2}n^{-1/2}(\rv)\frac{\partial n(\rv)}{\partial x}$, which gives
$\del \sqrt{n(\rv)} = \frac{1}{2} n^{-1/2}(\rv)\del n(\rv)$.}
The von Weizs\"acker term is also the exact kinetic energy for a spin-paired
two electron non-interacting system, as both particles occupy the same orbital.

One of the approximations available for $G[n]$, in an attempt to reproduce
the kinetic energy description of the KS method, relies on introducing
the Pauli kinetic energy functional $T_\theta [n]$. This functional is given as the
difference between the KS noninteracting kinetic energy functional and the von Weizs\"acker term,
$T_{\theta}[n] = T_s [n] - T_\text{W} [n]$ (Ref.~\onlinecite{Pauli_positivity}).
Including also the exchange-correlation functional $E_\text{xc}[n]$,
$G[n]$ is simply
\begin{align}
\label{G_levy}
G[n] = T_{\theta} [n] + E_\text{xc} [n].
\end{align}
Since the exact KS noninteracting kinetic energy functional involves explicit
orbitals, an orbital-free implementation of Eq.~(\ref{G_levy})
must rely on finding a suitable
approximation for $T_\theta [n]$ as a functional of the total density alone.
Several exact requirements (such as positivity) have already been
demonstrated for this term.~\cite{Pauli_positivity} Violation of these
conditions have also been associated with OFDFT functionals convergence
problems.~\cite{KTH06}

The ground-state density minimizing $E[n]$ is given by the variational principle 
under the constraint $\int \td \rv \, n(\rv) = N$, where $N$ is the number of
electrons in the system. Then $\frac{\delta E}{\delta n}$ from an Euler-Lagrange
equation is expanded, yielding
\begin{align}
\frac{\delta E}{\delta n} 
= & \frac{\delta n^{1/2}}{\delta n} \frac{\delta}{\delta n^{1/2}}
\left( n^{1/2} \left( -\frac{1}{2} \del^2 \right) n^{1/2} \right)
\nonumber \\
& + \int \td \rv \, n(\rv') \, |\rv - \rv'|^{-1} + v(\rv) + \frac{\delta G[n]}{\delta n}
\nonumber \\
= & n^{-1/2} \left( -\frac{1}{2} \del^2 \right) n^{1/2} + v_\text{eff}(\rv) = \mu,
\end{align}
where $\mu$ is the negative of the ionization energy,~\cite{orbital_free}
and the Hartree and $G[n]$ terms have been pushed into $v_{eff}(\rv)$.
Both sides are multiplied by $n^{1/2}$,  giving a Schr\"odinger-like equation:
\begin{align}
\label{ofeq}
\left( -\frac{1}{2} \del^2 + v_\text{eff}(\rv) \right) n^{1/2} = \mu \, n^{1/2},
\end{align}
where terms on the left side can be interpreted as the effective Hamiltonian.

One OFDFT energy functional we will reference in this work is the so-called TFD-$\lambda$W model.~\cite{orbital_free} It contains the Thomas-Fermi kinetic energy \cite{Thomas, Fermi}, the Dirac exchange \cite{Dirac}, and a fraction of the von Weizs\"acker kinetic energy functional. In terms of the previous partitioning~\eqref{G_levy}, the model is obtained by defining
\begin{align}
\label{G_fixed}
T_{\theta} [n] = & C_\text{F} \int \td \rv \, n^{5/3} 
\nonumber \\
& + \left(\lambda - 1 \right)
\int \td \rv \, n^{1/2} \left( -\frac{1}{2} \del \right) n^{1/2},
\\
E_\text{xc} [n] = & C_\text{x} \int \td \rv \, n^{4/3},
\end{align}
with $C_\text{F}=\frac{3}{10} (3 \pi^2)^{2/3}$ and $C_\text{x} = \frac{3}{4}
(\frac{3}{\pi})^{1/3}$. 
For the TFD-$\lambda$W model when $\lambda$ takes smaller values
such as $1/5$ or $1/9$, convergence problems appear and the von Weizs\"acker term in $T_{\theta}$ has been determined to be the root
of the problem (see Ref.~\onlinecite{Karasiev} for a review). Here we are removing those convergence problems by rearranging \eqref{ofeq} to
\begin{align*}
\left(-\frac{\lambda}{2}\del^2 + v_\text{eff}' \right)n^{1/2} = \mu \, n^{1/2},
\end{align*}
where the derivative of the von Weizs\"acker term in $T_{\theta}[n]$ is summed back to the kinetic energy operator. Then both sides are divided by $\lambda$ to get to equation
\begin{align}
\label{lambda_trick}
\left(-\frac{1}{2}\del^2 + \frac{1}{\lambda}{v'}_\text{eff} \right)
n^{1/2} = \frac{\mu}{\lambda} n^{1/2}.
\end{align}
The implementation is quite straightforward. Divide the potential
${v'}_\text{eff}$ by $\lambda$ after it is calculated. Then, after convergence 
has been achieved, multiply the eigenvalue $\frac{\mu}{\lambda}$ by $\lambda$
to get the correct eigenvalue and energies. This equation converged for $\lambda$ values
$1/5$ and $1/9$ for the systems shown in Table~\ref{tab:ae_atom}. 

\subsection{The PAW method}
\label{sec:OF_GPAW}
An attractive trait about the PAW method \cite{PAW_original} is that it is
a method that gives access to all-electron values. The method transforms the wave functions into pseudo wave functions with a smooth and therefore more convenient spacial
behavior. To obtain the all-electron wave function $\psi_n$ from a smooth wave function $\tilde{\psi_n}$ ($n$ is the state index), we seek a linear transformation
\begin{align}
\label{PAW_transformation}
\ket{\psi_n} = \hat{T}\ket{\tilde{\psi_n}}.
\end{align} 

The transformation operator is divided into atom-centered transformations which each has effect only inside one augmentation region. Atom-centered transformations are defined by atomic setups which contain the quantities related to a single atom, that is, smooth quantities for grid calculations and the unmodified atom quantities to define the transformation operator. Each element has to have its own atomic setup. Inside the augmentation sphere, the all-electron wave function can be expanded in atom-like partial waves basis $\{ \ket{\phi_i^a }\} $ ($i$ is a combination index for the principal, angular momentum, and magnetic quantum numbers respectively $(n,\ell,m)$).  The index $a$ indicates to which augmentation region the operator acts on. Then we define a set of auxiliary wave functions $\{ \ket{ \tilde{\phi}_i^a }\} $ that are smooth everywhere and
match the atom-like partial waves at the augmentation 
sphere's boundary. Finally, we define linear projector functions
$\bra{\tilde{p}_i^a}$ as zero outside the augmentation sphere and
satisfying the completeness relation
\begin{align*}
\sum_i \ket{\tilde{\phi_i^a}} \bra{\tilde{p}_i^a} = 1.
\end{align*}
In terms of the atom-like partial waves functions, the partial pseudo wave
functions, and the projectors, the transformation is written as
\begin{align*}
\hat{T} = 1 + \sum_{a,i} \left( \ket{\phi_i^a} - \ket{\tilde{\phi_i^a}}
\right) \bra{\tilde{p}_i^a}.
\end{align*}
The form of the Schr\"odinger equation within the PAW context is
\begin{align}
\label{PAW_equation}
\hat{\tilde{H}} \ket{\tilde{\psi_n}} = \varepsilon_n \hat{S} \ket{\tilde{\psi_n}},
\end{align}
where $\hat{\tilde{H}} = \hat{T}^{\dagger} \hat{H} \hat{T}$ is the transformed
Hamiltonian and $\hat{S}=\hat{T}^{\dagger} \hat{T}$ is the overlap operator. 
In summary, the all-electron wave-function, is obtained from the transformation \eqref{PAW_transformation} where the pseudo wave functions are obtained solving the eigenvalue equation \eqref{PAW_equation}. For details on combining PAW with the KSDFT method in the frozen core approximation see refs. \onlinecite{PAW_original,GPAW}.

The convenience of formulating equation \eqref{ofeq} in the PAW formalism is
two fold. First, PAW reproduces all-electron values offering freedom in adjusting the convergence parameters and the atomic setups allow 
tuning the numerical accuracy per element. Second, the PAW method
could provide answers to some of the convergence problems experienced by
other OFDFT implementations that make use of KS codes. We have implemented
the scheme in the open-source GPAW software,~\cite{GPAW} which is basically
PAW with real space grids and exchange and correlation functionals from 
LibXC (Ref.~\onlinecite{libxc}). Implementation in this setting gives various advantages.
GPAW itself is well written and community maintained with an active user base. 
A parallelization scaling almost ideally is available and we can use all of the functionals
provided by LibXC, including future updates. Implementation consists of
two major parts. First, we need to generate suitable orbital-free setups
and then ensure that they are loaded correctly to GPAW. Next, we force the
occupation of electronic bands so that only one band gets occupied.

\subsection{All-electron orbital-free implementation for PAW setup generation}
\label{sec:OF_KS}

\begin{table*}[t]
\caption{Energies of atoms calculated using the all-electron atomic code in
GPAW. The reference energies~\cite{Karasiev} are obtained also with an
all-electron code, a very small error should therefore be attained, as is the
case. The model is TFD-$\lambda$W, as defined in Eq.~\eqref{G_fixed}. All energies are in eV. The mean absolute error (MAE) is reported in the last row. For comparison, the LDA spin unpolarized KSDFT energies are also included.}
\label{tab:ae_atom}
\begin{ruledtabular}
\begin{tabular}{l | c c | c c | c c| c}
& E$_{\lambda=1}$ & E$_\text{ref}$ (Ref. ~\onlinecite{Karasiev}) & E$_{\lambda=1/5}$ & E$_\text{ref}$ (Ref. ~\onlinecite{Karasiev}) & E$_{\lambda=1/9}$ & E$_\text{ref}$ (Ref.~\onlinecite{Karasiev})  & E$_\text{LDA KS}$
\\
\hline 
H
& $-7.126$ & $-7.124$
& $-15.420$ & $-15.418$
& $-18.136$ & $-18.134$
& $-12.128$
\\
He
& $-40.205$ & $-40.205$
& $-76.693$ & $-76.693$
& $-87.699$ & $-87.697$
& $-77.130$
\\
Li
& $-111.716$ & $-111.714$ 
& $-199.262$ & $-199.261$
& $-224.535$ & $-224.535$
& $-199.588$
\\
Be
& $-231.086$ & $-231.085$
& $-394.133$ & $-394.133$
& $-439.820$ & $-439.821$
& $-393.112$
\\
B
& $-406.155$ & $-406.155$
& $-670.172$ & $-670.173$
& $-742.528$ & $-742.534$
& $-662.415$
\\
C
& $-643.738$ & $-643.737$
& $-1034.934$ & $-1034.936$
& $-1140.287$ & $-1140.302$
& $-1018.372$
\\
N
& $-949.908$ & $-949.906$
& $-1495.063$ & $-1495.073$
& $-1639.794$ & $-1639.821$
& $-1470.048$
\\
O
& $-1330.181$ & $-1330.180$
& $-2056.527$ & $-2056.542$
& $-2247.065$ & $-2247.114$
& $-2026.454$
\\
F
& $-1789.627$ & $-1789.628$
& $-2724.773$ & $-2724.802$
& $-2967.580$ & $-2967.658$
& $-2696.560$
\\
Ne
& $-2332.952$ & $-2332.953$
& $-3504.824$ & $-3504.871$
& $-3806.391$ & $-3806.512$
& $-3489.315$
\\
MAE
 & $0.001$ &
 & $0.011$ &
 & $0.030$ &
&
\end{tabular}

\end{ruledtabular}
\end{table*}
The first stage of the implementation is to generate suitable atomic setups for GPAW. The setups contain information of atomic quantities such as projector functions and partial waves, that were discussed in Section~\ref{sec:OF_GPAW}. The main quantities of interest for the atomic setups in this implementation are the partial waves and the cutoff value , which is the radius of augmentation sphere. In GPAW the partial waves are chosen as the solutions of the radial Kohn-Sham equation multiplied by spherical harmonics \cite{exact,GPAW}. Other quantities of setups, e.g. the projector functions, are generated from the partial wave and the cutoff value with standard GPAW routines \cite{exact,GPAW}, which make quantities as convergent as possible with respect to the chosen basis of partial waves. Smooth quantities for the calculations on the grid are generated to match the partial waves outside the augmentation sphere. Inside the augmentation sphere, the smooth quantities are just 6th order polynomials fitted at the boundary of the sphere. In the orbital-free setting each atomic setup then contains exactly one partial wave, the square root of density, which is obtained by solving Eq. \eqref{ofeq} for a single atom with the exchange-correlation term modified to introduce the Pauli kinetic term, Eq.~\eqref{G_levy}. Here, all the quantities, such as projector functions, have been generated as would be for a $(n,\ell)=(1,s)$ orbital, since we have a one-band system where the total occupation is set to the actual number of electrons.

Here lies one of the strengths of the GPAW implementation: we can change the
setup parameters per element. We need only study a few parameters:
the cutoff value $r_{c}$ of augmentation spheres and the projector functions.
The cutoff is simply chosen as the maximum radial value that allows the
all-electron energies to be obtained for a given desired grid spacing.
In the PAW formalism, we can choose the type of orbitals in the basis for
the expansion of the all-electron density. Our current choice 
uses a $1s$ orbital only, but it is possible to break the spherical symmetry
imposed by the use of the $1s$ orbital inside the augmentation sphere by
allowing the use of a higher number of bands or by simply changing the basis.

Table~\ref{tab:ae_atom} lists a comparison between total energy values
obtained with the atomic code and reference all-electron OFDFT calculations.
A small mean absolute error (MAE) is obtained for various values of $\lambda$ using
the TFD-$\lambda$W model (MAE of 0.001, 0.011, 0.030~eV for $\lambda=1, 1/5,
1/9$, respectively). For $\lambda$ different than one, the ground state of this 
model converges more easily when regrouping the energy terms differently (Eq.~\eqref{lambda_trick}).
Here the main point is to avoid having a potential term of the form
of the potential arising from the von Weizs\"acker term, since this will
produce a strong negative divergence in the Pauli potential for $\lambda<1$
(see the discussion in Ref.~\onlinecite{Karasiev}). When all such terms are
combined with the kinetic energy operator, the self-consistency cycle works
differently but nevertheless leads to convergence, which is faster thanks to
the somewhat smoother shape of the remaining terms in the Pauli potential.
We have to point out that the
above regrouping of terms used requires a higher number of grid points in the
all-electron calculation to reach the reported accuracy. The results shown
in Table~\ref{tab:ae_atom} were obtained with 1200 radial grid points. With 600
radial grid points, the MAE of the $\lambda=1$ model increases from 0.001 to 0.003~eV.

\subsection{OFDFT implementation using the GPAW method}

After generating the one-orbital setups, we must ensure that they are loaded
correctly. As setups are loaded to orbital-free GPAW, the number of valence
electrons is set to correspond to a situation where there is no frozen core
(valence equals the atomic number). Then, to actually solve the system, the
number of calculation bands is set to one, and the occupation number gives
the correct normalization. This is essentially the only modification needed
for the orbital-free calculations.

\subsubsection{Atoms and diatomic molecules}

\begin{table*}[t]
\caption{Comparison of binding energies ($\Delta E$ in eV) and bond lenghts ($r_e$ in \AA) of diatomic molecules for the TFD-$\lambda$W model with $\lambda = 1$ using all-electron (AE) and PAW method. The first two columns contains energies and bond lengths (in \AA) from an all-electron calculation.~\cite{molecules}. The next three columns contain GPAW results. First, we include PAW energies with bond lengths from the reference then results obtained with bond-length optimization in GPAW. In the last two columns, KSDFT with exchange-correlation LDA in a spin-polarized calculation values are also included.}
\begin{ruledtabular}
\begin{tabular}{l | cc ccc | cc}
& \multicolumn{2}{c}{AE OFDFT $\lambda=1$ (Ref. \onlinecite{molecules})}  & \multicolumn{3}{c}{PAW OFDFT $\lambda=1$} & \multicolumn{2}{c}{PAW KSDFT LDA (Ref. \onlinecite{PBE_ref_solid})}
\\
\hline
& $\Delta E$         & $r_e$          & $\Delta E$                  & $\Delta E_
\text{relaxed}$      & $r_e$       &$\Delta E$      & $r_e$
\\
H$_2$
& $-1.388$ & $1.730$
& $-1.385$ & $-1.385$
& $1.735 $ &-4.922 &      0.764 
\\
N$_2$
& $-12.599$ & $1.244$
& $-12.605$ & $-12.605$
& $1.242$ &  -11.539 &      1.095
\\
O$_2$
& $-13.606$ & $1.249$
& $-13.611$ & $-13.611$
& $1.251$ &  -7.550 &      1.204
\\
F$_2$
& $-14.476$ & $1.259$
& $-14.473$ & $-14.473$
& $1.262$ & -3.356 &      1.388
\\
HF
& $-4.000$ & $1.492$
& $-4.008$ & $-4.008$
& $1.491$ &  -7.047 &      0.931
\\
CO
& $-12.463$ & $1.238$
& $-12.450$ & $-12.453$
& $1.245$ &  -12.927 &      1.127
\\
MAE & & 
 & $0.006$
 & $0.006$
 & $0.003$  &&                                                                  
 
\end{tabular}

\end{ruledtabular}
\label{tab:gpaw_dimer}
\end{table*}

Results are obtained for atoms and diatomic molecules for the TFD-$\lambda$W
model with $\lambda = 1$. First, we tested that the all-electron values
presented in Table~\ref{tab:ae_atom} can be reproduced using PAW and real-space methods. For that purpose, we used the deviation from the all-electron value as the error to test the GPAW calculation parameters for non-periodic systems. We determined for the atoms that the errors obtained using PAW and real-space methods are of the order of the meV. Using the standard grid spacing of 0.18~{\AA}, a cutoff of 1.2~bohr and 1200 radial grid points for setup
generation we obtained a mean absolute error of 10~meV and the highest deviation for H is 20~meV. By decreasing the grid spacing to 0.14~{\AA}, for example, the mean absolute error decreases to 9~meV. In the following we use then setups generated with a cutoff of 1.2~bohr and 1200 radial grid points.

Using the derived setups, we calculated the binding energies of simple
diatomic molecules. The binding energy is calculated as the difference between the molecule and the atoms energy obtained in all cases with the  PAW method. The results obtained are listed in
Table~\ref{tab:gpaw_dimer}. As the table shows, the orbital-free GPAW results
agree with the reference all-electron calculation.~\cite{molecules}. The mean absolute error (MAE) of the order of meV agrees with the energy difference obtained between the atomic all-electron and the real-space PAW code for atoms. We optimized the bond length by
minimizing the forces acting on the atoms. In practice, we accomplished this
by using the quasi-Newton method provided by the Atomic Simulation
Environment~\cite{ASE} (ASE). The huge number
of iterations required to converge the orbital-free scheme in the atomic code
is reduced. In the orbital-free GPAW approach, the number of iterations
required by calculations are of the same order of magnitude as the normal
KS calculations. Thus, using PAW in combination with real-space methods is a promising
route for OFDFT calculations. Now, let us study the OFDFT energies of bulk
Li and diamond as representatives of periodic systems.

\subsubsection{Periodic solids}

\begin{table}[t]
\caption{Lattice constant ($a_o$ in \AA) and bulk modulus ($B_0$ in GPa) of bulk Li bcc as a function of different methods and functionals. OFDFT calculations use the same energy functional: the Pauli kinetic term is the same as in the TFD-$\lambda$W model with $\lambda=1$, Eq.~\eqref{G_fixed}, and the exchange-correlation functional $E_\text{xc} [n] = E_\text{x}^\text{PBE} [n]$ is the PBE exchange.~\cite{PBE} The first row present OFDFT calculations using PAW while the second and third present OFDFT calculations with pseudopotentials (PP) that are obtained from reference work.~\cite{Karasiev}. In Ref. \onlinecite{Karasiev}, PP spd1 and mod1 refer to different ways to construct local pseudopotentials (the first one uses a normalized linear combination and the second one uses a direct modification of the effective potential). The values from PAW KS DFT method with PBE exchange-correlation functional are included in the last row.}
\begin{ruledtabular}
\begin{tabular}{llll|c|c}
Method& & & &$a_0$ & $B_0$ \\
\hline
PAW &OFDFT &$\lambda=1$& & 1.646 & 945 \\
PPspd1 &OFDFT &$\lambda=1$ &Ref. \onlinecite{Karasiev}& 3.43 & 15.2 \\
PPmod1 &OFDFT &$\lambda=1$ &Ref. \onlinecite{Karasiev}& 3.43 & 14.9 \\
\hline
PAW &KSDFT &PBE &Ref. \onlinecite{PBE_ref_solid}&3.435 & 14.0
\end{tabular}

\end{ruledtabular}
\label{tab:gpaw_Li_results}
\end{table}

We performed the OFDFT calculations for Li in a body-centered cubic (bcc)
configuration and for carbon in a diamond configuration. For Li first, we
used a conventional unit cell with two atoms and to check the effect of size, we 
repeated the calculations with a cubic unit cell containing 16 atoms.
Bulk modulus simulation with PAW and real-space methods usually requires a small
grid spacing. A small strain (close to 1~\% of the lattice constant)
is applied with constant number of grid points and convergence is verified by
increasing the number of grid points (equivalently decreasing the grid spacing).
For carbon and lithium, we used 20 grid points along each direction
(corresponding to a grid spacing around $h = 0.1 \text{~\AA}$) and cell increments of 0.02~{\AA}. 
For the cutoff in the setup generation, we used 1.0~bohr for Li and 1.2~bohr for C. With this cutoff Li and C bond distances of 2.0 and 2.4~bohr respectively (1.05 and  1.27~\AA) can be calculated without inducing errors coming from an augmentation sphere superposition \cite{KST12}. The values in Tables~\ref{tab:gpaw_Li_results} and \ref{tab:gpaw_diamond}  have therefore been calculated without augmentation sphere superposition problems.

Table~\ref{tab:gpaw_Li_results} summarizes Li results.
%Refs.~\onlinecite{Karasiev,Huang08}) differ considerably from the bare
The Li bulk modulus and lattice constant using PAW are extremely different from the reported pseudopotential values. Indeed, such pseudopotentials are obtained by fitting procedures to match KS or experimental values and the energies are then calculated with a OFDFT energy functional. In contrast, in the present PAW calculation the setups and the energy have been calculated using the same OFDFT energy functional and the energy obtained is an all-electron energy \cite{PAW_original}. Thus from this comparison for Li, we clearly conclude that in order to assess the performance of OFDFT functionals, it is necessary to use all-electron values, and the PAW method is for now an excellent candidate,
as shown throughout this work.

\begin{table}[t]
\caption{Bulk diamond lattice constant ($a_0$ in \AA), bulk modulus ($B_0$ in GPa) as a function of different methods and functionals using PAW. In the OFDFT functional, the Pauli kinetic term used is the same as in the TFD-$\lambda$W model with $\lambda=1$, Eq.~\eqref{G_fixed}, multiplied by a scaling parameter $\gamma$ and the exchange-correlation functional $E_\text{xc} [n] = E_\text{x}^\text{PBE} [n]$ is the PBE exchange.~\cite{PBE}. The KSDFT with PBE exchange-correlation and experimental values are from Ref. \onlinecite{PBE_ref_solid}.}
\begin{ruledtabular}
\begin{tabular}{lll| c | c }
Method && &$a_0$ & $B_0$ \\
%& Cohesive energy\\
\hline
OFDFT &$\gamma=1.00$&
& 2.815 & 1407 
%& 24.5 
\\ 
%OFDFT &$\gamma=1.05$&
%& 3.011 & 1016 
%& 21.1
% \\ 
OFDFT &$\gamma=1.10$&
& 3.213 & 741 
%& 18.3 
\\ 
%OFDFT &$\gamma=1.15$&
%& 3.419 & 535 
%& 15.9 
%\\ 
%OFDFT &$\gamma=1.20$&
%& 3.629 & 411 
%& 13.8 
%\\ 
OFDFT &$\gamma=1.1859$&
& 3.570 & 446 
%& 14.4 
\\
\hline 
Exp. & &Ref. \onlinecite{PBE_ref_solid}& 3.567 & 443 
%& 7.58 
\\
KSDFT &PBE &Ref. \onlinecite{PBE_ref_solid}& 3.575 & 431 
%& 
\\ 
\end{tabular}

\end{ruledtabular}
\label{tab:gpaw_diamond}
\end{table}

By giving access to all-electron values, the present implementation opens
the way to test the accuracy of OFDFT energy functionals. As an example of
such studies, we have tested a simple parametrization using as reference
bulk diamond. The model contains as Pauli kinetic term $T_\text{F}$ the Pauli term of the TFD-$\lambda$W model with $\lambda=1$, Eq.~\eqref{G_fixed}, multiplied by a scaling parameter $\gamma$. In other words, the Pauli kinetic term is a parametrized Thomas-Fermi kinetic functional \cite{Thomas,Fermi}. The exchange-correlation functional $E_\text{xc} [n] = E_\text{x}^\text{PBE}$ is the PBE exchange.~\cite{PBE}. Such a kinetic decomposition fulfills the exact properties of the Pauli kinetic energy functional.~\cite{Pauli_positivity}
We determined that increasing the fraction of the Fermi kinetic functional
present increases the lattice constant in a smooth way. As expected from such
behavior, the increased $\gamma$ also decreases the cohesive energy and
decreases the bulk modulus almost linearly. Finally, we should note that it
is possible to fit the parameter $\gamma$ to reproduce experimental values
for the lattice parameter. With the numerically obtained $\gamma=1.1859$,
we performed a self-consistent evaluation of lattice constant and bulk modulus. 
The values are reported in Table~\ref{tab:gpaw_diamond}, and with respect to 
experimental values, they deviate by 0.1\% and 0.7(\% for lattice parameter
and bulk modulus, respectively). More importantly, this value is comparable
to the accuracy reached by the PBE functionals in the KS method (the reference KS PBE lattice constant and bulk modulus are 3.575~{\AA} and 431~GPa, see
Ref.~\onlinecite{PBE_ref_solid} for a calculation using PAW).
The good value for the bulk modulus indicates that the evolution of the
system's energy as a function of interatomic distances is basically well
described around the equilibrium lattice constant. We computed however a large cohesive energy: 14.4 eV for fitted $\gamma$ vs 7.58 exp.\cite{Kittel}. The large cohesive energy 
means that the drop in energy of the overall system as the atoms are brought
together to form the crystal is too large. Since the change in energy as
the atoms are moved closer/apart around equilibrium is well described this
probably means the reference energy for the isolated atoms is too high.
For bulk lithium, the variation of the bulk modulus and 
lattice constant with increased $\gamma$ is similar to the diamond case.
Using the same basic functionals, increasing the amount of kinetic Thomas-Fermi functional via 
increasing $\gamma$, decreases the bulk modulus and increases the lattice parameter.  
A fitted $\gamma=1.7996$ gives a Li equilibrium 
lattice parameter of 3.476 \AA~and a bulk modulus of 28 GPa. The bulk lithium equilibrium lattice parameter can be 
approached very accurately but the associated bulk modulus while being the right order of magnitude is twice the experimental value. 
For diamond and bulk Li the procedure of fitting $\gamma$ to reproduce the experimental lattice parameters led to different values of $\gamma$ in each case. In other words, this very simple fitted functional for diamond is not transferable to bulk Li. To improve the accuracy and transferability of OFDFT functionals, even more terms in the kinetic expansion, better exchange functionals, and correlation functionals can be added. Strategies of fitting parameters can be improved by including a larger set of systems and properties.

\subsubsection{Performance}

\begin{figure}[t]
\centering
\includegraphics[angle=0,width=\columnwidth]{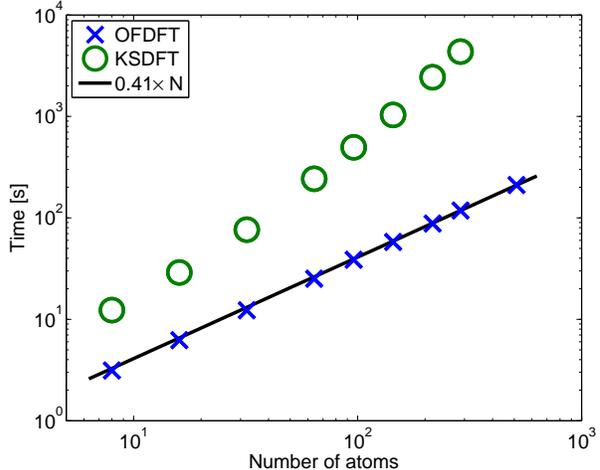}
\caption{Total computational time for a fixed number of self-consistent
iterations versus the number of atoms in the unit cell for diamond using PAW and real-space methods. Note that a logarithmic scale is used. The KS DFT calculation uses the PBE exchange-correlation functional. OFDFT is computed with the
TFD-$\lambda$W model with $\lambda=1.188$.}
\label{fig:time}
\end{figure}

Finally, we include in this section a comparison of the KS DFT and OFDFT
performances. Because OFDFT is implemented reusing the KS DFT method, all
algorithm prefactors are the same and therefore a good comparison can be
carried out. For the KS DFT method, when increasing the size of the system
the time scaling becomes inevitably cubic. OFDFT implemented with the same
real space code should retain a linear time scaling for all sizes. 
To illustrate both scaling behaviors, we present in Fig.~\ref{fig:time}
the total time (including initialization) for computing a small number of
steps. The time needed for completing four iterations of the self-consistent
cycle is shown as a function of the number of atoms in a diamond unit cell
of variable size.  When the system size increases, the orthogonalization algorithm --which displays
cubic scaling-- determines the total time of a KS DFT cycle, while OFDFT
remains scaling linearly with size. When increasing the size of the diamond
cell it was found that convergence was not reached simply reusing the
parameters of the conjugate gradient algorithm employed. A more detailed
study of the algorithm parameters is needed before simulations
of large (meaning millions) of atoms can be attained.

The number of iterations to reach self-consistency with default KS
convergence parameters is another important consideration. Convergence of
the all-electron atomic method, used to generate the setups, requires
typically 10 to 100 more iterations for OFDFT than for KS DFT,
consistent with previous findings.~\cite{Karasiev}
The large number of self-consistent iterations needed to reach convergence in the atomic code is related to the density mixing parameter used in the Pulay density mixing algorithm \cite{Pulay}. The mixing parameter controls how the density is updated in the self-consistency cycle, with small mixing parameter giving slow mixing and helping to the overall convergence stability. The small parameter in the  density mixing only needs to be used once per atom inside the setup generation (and does not scale with size), therefore no further exploration was performed. Using PAW and real-space methods stabilizes the self-consistency
iteration dramatically. One only needs tens of iterations for atoms,
dimers and solids for the TFD-$\lambda$W mode, see Table~\ref{tab:iterations}.
This is comparable to the number of iterations typically needed by
\textit{ab initio} functionals.

\begin{table}[b]
\caption{Number of iterations to reach default convergence in the
self-consistent cycle using PAW and real-space methods. The OFDFT functional is the TFD-$\lambda$W model with $\lambda=1$ and the KS exchange-correlation functional is LDA. The grid spacing $h$ is in \AA.}
\begin{ruledtabular}
\begin{tabular}{l|cc|cc}
& \multicolumn{2}{c}{OFDFT}& \multicolumn{2}{c}{KSDFT}
\\
\hline
&  $h$ & Iterations &$h$&Iterations\\
%\hline 
H                      & 0.14                  & 18 & 0.18 & 11 
\\
N$_2$                  & 0.18                 & 32  &0.18 & 13
\\
Li bcc                  & 0.025               & 27 & 0.0771 & 10
\end{tabular}

\end{ruledtabular}
\label{tab:iterations}
\end{table}

\section{Conclusion}
This work demonstrates that the PAW method, in combination with KS codes,
can reproduce all-electron OFDFT energies of atoms and dimers without
convergence problems. Indeed, the PAW calculation converges in a number
of self-consistent steps comparable to the KS method. This combination of
methods offers a promising route for reaching easy-to-converge large-scale
OFDFT calculations. Moreover, when testing the accuracy of OFDFT kinetic functionals, a method such as the one presented here becomes necessary.

Before this work can be extended to large-scale simulations, the exact
relation between the setup cutoff and the resulting energy accuracy for
real-space parameters must be studied for the elements involved. It is important
to determine the minimum number of self-consistent iterations that can be
reached once the target accuracy is defined. With the present knowledge, an
increase in size of the same system requires fine tuning of simulation
parameters to achieve convergence.   

Interestingly, with the simplest kinetic approximation for diamond, we
obtained a lattice constant and bulk modulus comparable to the PAW KS PBE values.
This begs the question, then, to what extent such simplicity can be
maintained when choosing the building blocks of more accurate and transferable kinetic functionals. We hope this work will contribute to the development of better OFDFT functionals by allowing access to all-electron energy values of
periodic and non-periodic systems.

\begin{acknowledgments}
This work was supported by Academy of Finland projects 279240 and 251748.
\end{acknowledgments}
\end{document}